\documentclass[11pt]{article}

\def\d{{\rm d}}
\def\cH{{\cal H}}
\def\cW{{\cal W}}
\def\ve{\varepsilon}
\newcommand{\eqn}[1]{(\ref{#1})}
\def\beq{\begin{equation}}
\def\eeq{\end{equation}}
\def\Id{1\hspace{-4.5pt}1}
\def\z{\zeta}
\def\zb{\bar\zeta}
\def\Vir{{\sf Vir}}
\def\Zbb{{\sf Z\hspace{-4.5pt}Z}}
\def\Rbb{{\sf I\hspace{-1pt}R}}

\begin{document}

\vskip 0.5cm
\begin{flushright}
KCL-MTH-02-01\\
PAR-LPTHE-02-07\\
{\tt hep-th/0201231}\\
\end{flushright}
\vskip 1.cm
\begin{center}
{\Large\bf A non-rational CFT with central charge 1}\\
\vskip 1.3cm
I.~Runkel${}^1$ and G.M.T.~Watts${}^2$\\
\vskip 0.6cm
${}^1${\sl LPTHE - Univ.~Paris VI, 4 place Jussieu, F -- 75252 Paris} \\
{\tt ingo@lpthe.jussieu.fr} \\[5pt]
${}^1${\sl Dept.~of Mathematics, King's College, The Strand, GB -- London WC2R 2LS}\\
{\tt gmtw@mth.kcl.ac.uk}
\end{center}

\vskip 0.9cm
\begin{abstract}
\vskip0.15cm
\noindent
  Two dimensional conformal field theories with central charge
  one are discussed. After a short review of theories based
  on one free boson, a different CFT is described, 
  which is obtained as a limit of minimal models.
\end{abstract}

%%%%%%%%%%%%%%%%%%%%%%%%%%%%%%%%%%%%%%%%%%%%%%%%%%%%
\setcounter{footnote}{0}
\def\thefootnote{\fnsymbol{footnote}}

%\section{Introduction}

We will be concerned with unitary two dimensional conformal field
theories. Such theories describe, for example, critical points of
equilibrium statistical mechanics systems 
in two dimensions, or vacua of
string theory around which a perturbative expansion of string
amplitudes can be computed.

Since the discovery by Belavin, Polyakov and Zamolodchikov of
a discrete series of CFTs, the minimal models \cite{BPZ}, much effort
has been spent in trying to classify all CFTs. An important parameter
in this problem is the {\em central charge} $c$ of a CFT, which appears in
the two point function of the stress tensor $T(\z)$ on the
complex plane
\beq
  \langle T(\z) T(\z') \rangle = \frac{c/2}{(\z-\z')^4}\;.
\eeq
Another useful notion is that of a {\em rational} CFT.
Roughly speaking, a CFT is rational when the algebra
generated by all conserved charges is large enough
for the Hilbert space to decompose into only finitely many
representations.

Here we are only considering unitary 
CFTs which makes the classification problem somewhat
easier, as for example logarithmic CFTs \cite{lcft}
cannot occur.
Nonetheless it turned out to be a
very difficult problem. A very brief summary of the state of affairs
could take the following form
\begin{itemize}
\item $c<1$: All unitary CFTs with central charge less than one are
known, they are the minimal models of \cite{BPZ} whose different 
modular invariants were given in \cite{CIZ}. 
Their central charges take the values
\beq
  c = 1 - \frac{6}{p(p+1)} \;.
\eeq
with $p=2,3,4,\dots$. For $p=2$ on has $c=0$ and the theory is
trivial in the sense that its only state is the vacuum. 
$p=3$ gives $c=1/2$ and one obtains the Ising model. For
$p\rightarrow\infty$ the central charge approaches one.
All these theories
are rational CFTs.
\item $c=1$: All rational CFTs with central charge one are 
believed to be known 
\cite{Gins,Kir}, but no complete list also including all 
non-rational\footnote{
  It is customary to insert the notion of ``quasi-rational'' 
  as another step between
  rational and non-rational. For the present purposes we can think
  of non-rational as ``not rational''; what is said remains true.}  
theories exists. Until recently all examples of 
non-rational CFTs with $c{=}1$ were
based on the free boson. Here 
we want to describe another example of such a theory \cite{RW}.
\item $c>1$: Here the situation is even more hopeless. The list
of known CFTs with central charge greater than one is enormous
and includes for example WZW models and cosets thereof, supersymmetric
theories, theories with $\cW$--symmetry or Liouville theory
\cite{Ybk}. 
The question of a 
complete classification seems quite out of reach, the hardest part
is certainly to get a handle on the non-rational theories.
\end{itemize}
The value $c=1$ is special, because it is the smallest central charge
where we do not have complete knowledge, and it is also the smallest
$c$ for which non-rational theories can occur. From hereon
we concentrate on models with $c=1$, starting with the free boson.

\section*{Free boson}

In \cite{Gins} Ginsparg presented a list of theories with central
charge one based on the free boson. The resulting moduli space
is drawn below
\vspace{-4mm}
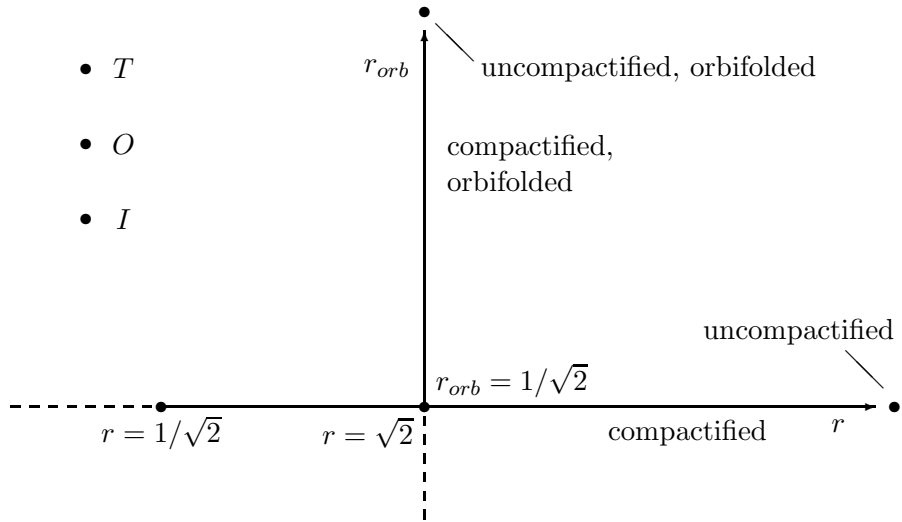
\begin{center}
\setlength{\unitlength}{0.5mm}
\begin{picture}(260,160)(0,0)
% horizontal axis
\put(10,40){\line(1,0){2.5}}
\put(15,40){\line(1,0){2.5}}
\put(20,40){\line(1,0){2.5}}
\put(25,40){\line(1,0){2.5}}
\put(30,40){\line(1,0){2.5}}
\put(35,40){\line(1,0){2.5}}
\put(40,40){\line(1,0){2.5}}
\put(45,40){\line(1,0){2.5}}
\put(50,40){\vector(1,0){190}}
% vertical axis
\put(120,40){\vector(0,1){100}}
\put(120,10){\line(0,1){2.5}}
\put(120,15){\line(0,1){2.5}}
\put(120,20){\line(0,1){2.5}}
\put(120,25){\line(0,1){2.5}}
\put(120,30){\line(0,1){2.5}}
\put(120,35){\line(0,1){2.5}}
% special points
\put(50,40){\circle*{3}} % self dual comp
\put(120,40){\circle*{3}} % self dual orb
\put(120,145){\circle*{3}} % uncomp orb
\put(245,40){\circle*{3}} % uncomp
\put(30,130){\circle*{3}} % T
\put(30,110){\circle*{3}} % O
\put(30, 90){\circle*{3}} % I
% text
\put(40,130){\makebox(0,0)[cc]{$T$}}
\put(40,110){\makebox(0,0)[cc]{$O$}}
\put(40, 90){\makebox(0,0)[cc]{$I$}}
\put(230,35){\makebox(0,0)[cc]{$r$}}
\put(110,130){\makebox(0,0)[cc]{$r_{orb}$}}
\put(50,33){\makebox(0,0)[cc]{$r=1/\sqrt{2}$}}
\put(123,47){\makebox(0,0)[lc]{$r_{orb}=1/\sqrt{2}$}}
\put(117,33){\makebox(0,0)[rc]{$r=\sqrt{2}$}}
\put(126,109){\makebox(0,0)[lc]{compactified,}}
\put(126,100){\makebox(0,0)[lc]{orbifolded}}
\put(190,33){\makebox(0,0)[cc]{compactified}}
\put(245,60){\makebox(0,0)[rc]{uncompactified}}
  \put(242,45){\line(-1,1){10}}
\put(135,130){\makebox(0,0)[lc]{uncompactified, orbifolded}}
  \put(123,142){\line(1,-1){10}}
\end{picture}
\end{center}
\vspace{-4mm}
The horizontal axis corresponds to models where the target space of
the free boson is periodically identified, $X \sim X + 2\pi r$. 
T--duality says that the free boson compactified on radius $r$ is not
different from the free boson on radius $1/2r$ (in the conventions
of \cite{Gins}). This is the reason the horizontal line ends at
$r=1/\sqrt{2}$. On the right end we have included a point which
corresponds to the uncompactified free boson, i.e.\ to infinite
radius. 

The vertical line is the moduli space of orbifold models. Those arise
when the free boson is modded out by the symmetry $X\rightarrow -X$ of
the action. In this procedure the Hilbert space is projected onto
invariant states and twisted sectors (corresponding to strings that
close up to an action of the symmetry) are added to the Hilbert space.

The orbifolded theory at $r=1/\sqrt{2}$ and the unorbifolded theory
at $r=\sqrt{2}$ turn out to be equivalent (in particular their
partition functions are equal), and hence the two lines meet at that
point. 

Ultimately this is related to an enhancement of the symmetry of the
free boson compactified at the self-dual radius $r=1/\sqrt{2}$, 
where it is equal to the WZW model $su(2)_1$. This also accounts for
the three isolated points, which are obtained by modding out the
$su(2)$ theory by the tetrahedral, octahedral and icosahedral
subgroups. 

The question is now if the moduli space of $c=1$ theories based on the
free boson exhausts already all unitary CFTs with $c=1$. It has been argued
\cite{Kir} that this list does in fact include all {\em rational}
theories with $c=1$. The rational theories are those, for which 
the radius squares to a rational number, in these cases additional
conserved charges appear.
As we will see in a moment, there is however at
least one {\em non-rational}
theory which is not included in this list.

\section*{Limit of minimal models}

To work out the quantum theory of the free boson, one usually uses
canonical quantisation starting from the classical action. Here we
would like not to worry whether a CFT can be obtained 
from an action and rather define it directly in terms of 
the following data:
\begin{itemize}
 \item the Hilbert space $\cH$ of the theory.
 \item the two-- and three--point functions on the complex plane.
 \item the boundary state for the unit disc.
\end{itemize}
To see why this is enough, consider for example a four point function
on the complex plane. By inserting a complete basis of states, it can
be reduced to an integral over three point functions as in
\beq
  \langle \phi(x_1)\phi(x_2)\phi(x_3)\phi(x_4) \rangle
  = \int \!\d p\; \langle \phi(x_1)\phi(x_2) \,|p\rangle
  \langle p| \,\phi(x_3)\phi(x_4) \rangle \;.
  \label{basis}
\eeq
The boundary state is needed if one wants to evaluate
correlators on surfaces
with a boundary (via a conformal mapping to the unit disc).

The model we want to
describe is obtained as a limit of minimal models $M_p$. To be more
precise, for an allowed value of the central charge $c<1$, there can be
one, two or three distinct CFTs with different modular invariant
partition functions \cite{CIZ}. $M_p$ denotes the simplest, the
charge conjugation modular invariant.

The limit we are interested in is that of $p\rightarrow\infty$, which
leads to $c\rightarrow 1$. The minimal models $M_p$ are known in
great detail, in particular there are explicit expressions for the
OPE structure constants \cite{DF}. Using these, one can formulate a 
limiting theory $M_\infty$ in terms of
the three bits of data described above,
which has central charge one. The details of that construction can be
found in \cite{RW}, here we just present the outcome. The explicit
formulas are given mainly to illustrate that everything fits on one
page. 

The Hilbert space of $M_\infty$ decomposes into representations 
of ${\rm \Vir}{\otimes}{\rm \Vir}$, where $\Vir$ is the Virasoro
algebra, the algebra of modes of the holomorphic 
(resp.\ anti-holomorphic) component of the stress tensor. 
A representation of $\Vir$ is characterised by the central charge
(which is one) and the lowest conformal weight $h$ in the 
representation. Paradoxically, the state with lowest conformal weight
is usually called a
``highest weight state'', in accordance with
representation theory of Lie algebras, and we will also stick to that
name. 

The Hilbert space $\cH$ consists of highest weight states
$|x\rangle$ and the ${\rm \Vir}{\otimes}{\rm \Vir}$ modules build on
them. The parameter $x$ takes any value in the set
$\Rbb_{>0} - \Zbb_{>0}$ and denotes a state with left/right
conformal weight equal to $h_x=x^2/4$.
A field $\phi_x$ corresponding to a highest weight state
$|x\rangle$ in $\cH$ is called a
{\em primary field}.

The two and three point functions of primary fields 
$\phi_x, \phi_y, \phi_z$ read
\beq
  \langle\;\phi_x(\z_1,\zb_1)\,\phi_y(\z_2,\zb_2)\;\rangle = 
    \delta(x{-}y) \cdot |\z_1-\z_2|^{-x^2} \;, 
\eeq
\beq\begin{array}{rl}
  \langle\;\phi_x(\z_1,\zb_1)\,
    \phi_y(\z_2,\zb_2)\,\phi_z(\z_3,\zb_3)\;\rangle &= 
    c(x,y,z) \cdot |\z_{12}|^{(z^2-x^2-y^2)/2} \\
    & \qquad \times |\z_{13}|^{(y^2-x^2-z^2)/2}\;
    |\z_{23}|^{(x^2-y^2-z^2)/2} \;,
\end{array}\eeq
where $\z_{ij}=\z_i-\z_j$ and
\[
  c(x,y,z) = P(x,y,z) \cdot \exp(Q(x,y,z)) 
\]
\[
  P(x,y,z) = \Bigg\{
  \begin{array}{ll}
  \frac{1}{2} : &\big(\; [x]{+}[y]{+}[z] {\rm~even,~and~} 
  |f_x{-}f_y|<f_z<\min(f_x{+}f_y\,,\,2{-}f_x{-}f_y) \;\big)\;\;{\rm~or}\\
  & \big(\; [x]{+}[y]{+}[z] {\rm~odd,~and~} 
  |f_x{-}f_y|<1{-}f_z<\min(f_x{+}f_y\,,\,2{-}f_x{-}f_y) \;\big)\\
  0 : & {\rm otherwise} \end{array} 
\]
\[
  Q(x,y,z) = \int_0^1 \hspace{-5pt}
  \frac{\d\beta}{({-}\ln\beta)\cdot(1{-}\beta)^2} \cdot \Big\{ \;
  2+\sum_{\ve=\pm 1}\big(\beta^{\ve x}{+}\beta^{\ve y}{+}
  \beta^{\ve z} \big) -
  \hspace{-10pt}\sum_{\ve_x,\ve_y,\ve_z=\pm 1} \hspace{-10pt}
  \beta^{(\ve_x x+\ve_y y+\ve_z z)/2}
  \Big\}
\]
Here $[x]$ denotes the largest integer
less than
or equal to $x$ and $f_x=x-[x]$ is the fractional
part of $x$. 

Finally, the boundary states which are obtained from the limit of
minimal models are
\beq
  \langle a| =  2^{3/4} \int_0^\infty \!\d x\,
  (-1)^{[x]} \sin(a\pi x)\, \langle\!\langle x | \;,
\eeq
where $a\in\Zbb_{>0}$ and $\langle\!\langle x |$ is the 
Ishibashi state in the representation with highest weight state
$\langle x |$. These boundary conditions and their boundary field
content are discussed in detail in \cite{GRW}.

In analogy with Liouville theory \cite{fzz} one would
expect that there are more conformal boundary conditions than these,
this remains for future work. In fact the expressions for
the two and three point functions above also look very similar to
Liouville theory.
This raises the interesting question 
of whether or not
$M_\infty$ can be obtained as a limiting form of 
Liouville theory, where $c=1$ is approached from above. 

Let us briefly discuss two subtleties of the model $M_\infty$ which
are also encountered in Liouville theory \cite{te}. 

The first subtlety is that the Hilbert space $\cH$ of $M_\infty$ does
not contain a state of conformal weight zero, i.e.\ there is no vacuum
state in $\cH$. In rational CFT one has a strict state--field
correspondence, i.e.\ for every field in the theory there is a state
in the Hilbert space. In non-rational theories this need no longer be
the case. In $M_\infty$ we can use the Ward identities to
%consistently 
define correlation functions of the identity $\Id$ and
the stress tensor $T(\z)$ consistently,
even though the Hilbert space does not
contain states corresponding to these fields. E.g.\ trivially
$\langle\; \phi \cdots \phi \;\Id \;\rangle
= \langle\; \phi \cdots \phi \;\rangle$ or 
\beq
 \langle\; \phi \cdots \phi \;T(\z) \;\rangle  
 = \sum_k \Big(\, \frac{h_k}{(\z-\z_k)^2}+\frac{1}{\z-\z_k} 
 \frac{\partial}{\partial \z_k} \,\Big) 
 \langle\; \phi \cdots \phi \;\rangle
\eeq
The important property of the Hilbert space is that it is complete in
the sense that there is a vector for every possible state of the
system. The completeness allows us to insert a basis of states as
in \eqn{basis}. 

Another subtlety arises from the OPE. Let $\phi_x$ and $\phi_y$ be two
primary fields. Their OPE would take the form
\beq
  \phi_x \, \phi_y = \int \!\d z\; c(x,y,z) ( \phi_z + {\rm descendants} )\;,
  \label{wrongope}
\eeq
where the integral is over $\Rbb_{>0}-\Zbb_{>0}$ and we have suppressed
the coordinate dependence in that expression. On the other hand the
two point function was $\langle\phi_x\phi_y\rangle = \delta(x{-}y)$. If
we were to
insert the OPE \eqn{wrongope} into the two point function we would
obtain zero, since only a field of weight zero can have a nonzero
one-point function and there is no such field in $\cH$. 

The way out of this problem is to think of the fields $\phi_x$ as
distributions and to work with the analogue 
of test functions instead, 
i.e.\ with ``smeared fields''
\beq
  \phi_f(\z,\zb) = \int \!\d x\; f(x)\, \phi_x(\z,\zb) \;.
  \label{smear}
\eeq
Note that the fields are smeared in representation space, not over
position. Further we can formally define 
$|0\rangle = \lim_{x\rightarrow 0} x^{-1} |x\rangle$ and think of 
$|0\rangle$ as shorthand notation for this limit. In particular it is
not a state in $\cH$. The OPE can now be formulated in terms of
smeared fields and inserting this OPE in the two point function gives
\beq
  \langle 0 |\,\phi_f \,\phi_g\,|0\rangle = \int \!\d x\; f(x)\, g(x) \;,
\eeq
where the position dependence has again been suppressed. This is
precisely the result one obtains when inserting the definition
\eqn{smear} in the two point function
$\langle\phi_x\phi_y\rangle = \delta(x-y)$.

\section*{Open questions and conclusion}

We have defined the CFT $M_\infty$ with central charge one as a limit of
minimal models. While its partition function is proportional to that of 
an uncompactified free boson, its operator algebra is entirely
different. In $M_\infty$, the OPE of any two fields contains
continuous intervals of primary fields, whereas the generic OPE of the
free boson contains only one primary field (two for the orbifolded
free boson). The exception is the OPE of two twist fields in the
uncompactified, orbifolded free boson, which contains all untwisted
representations, but otherwise the operator algebra 
(and the partition function) is different also in this case.

Thus the question remains what the interpretation of the 
quantum field theory $M_\infty$ should be. We do not know the answer
to that, but let us make two suggestive remarks.

First note that there is a
qualitative Landau-Ginzburg description of the minimal model $M_p$
\cite{z} which consists of one free
scalar field $X(\z,\zb)$ perturbed by a potential 
$g X^{2(p{-}1)}$.
Taking $p$ to infinity means that the
potential approaches the form of a square well with walls at 
$X=\pm 1$, forcing $X$ to take values only in the interval
$[-1,1]$. The resulting action is that of a membrane fluctuating
freely between two walls\footnote{
  The Landau Ginzburg interpretation was pointed out by J.~Cardy.}.
However since the Landau Ginzburg action is not yet conformal, one has to 
still find the end point of the renormalisation group flow, and it is
not clear that the limit of minimal models leading to 
$M_\infty$ has
anything to do with the limit of the potentials in the classical
action.
This has been investigated further
in the context of a {\em different} limit of the minimal models $M_p$
\cite{RWe} to that we study.

The second remark concerns the target space interpretation of coset
WZW models. The minimal model $M_p$ can be written as the coset
$su(2)_{p-2} \oplus su(2)_1 / su(2)_{p-1}$. In the limit 
$p\rightarrow\infty$ the target space should approach the 
quotient of classical
group manifold. If we interpret the factor $su(2)_1$ as some internal
degree of freedom, we are left with $su(2)/su(2)$ as target space,
where the quotient $su(2)$ acts by conjugation. 
$su(2)$ has the topology of an $S^3$ and the conjugacy classes
are $S^2$'s, so that the quotient space has the topology of an
interval\footnote{
  The target space interpretation for minimal models seen as
  cosets was pointed out to us by V.~Schomerus.}. 

Both points indicate that the theory might have something to do with a
sigma model taking values in an interval, but any detailed connection
remains to be worked out. In particular there has to be an effect
which allows the spectrum to be continuous, even though the target
space is compact.

Other interesting open problems are the precise connection 
between $M_\infty$ and Liouville theory and whether the calculation can
be repeated for other series with a limit point in the central charge,
such as super minimal models, WZW models and cosets thereof. This might
also help to find the correct interpretation for $M_\infty$. It would
also be worth investigating if theories obtained in this way can be
used as string theory vacua.

%%%%%%%%%%%%%%%%%%%%%%

{\bf Acknowledgments} -- This work has been presented by IR at
the RTN meeting ``The quantum structure of space-time and the geometric 
nature of the fundamental interactions''. The authors would like to thank
T.~Gannon and C.~Schweigert
for helpful comments.

%%%%%%%%%%%%%%%%%%%%%%

\end{document}